\documentstyle[12pt]{article}
\begin{document}
\baselineskip 14.5pt 
\parindent=1cm
\parskip 3mm

\title{Massive gluons and quarks and the equation of state \\
       obtained from SU(3) lattice QCD }

\author{ P\'eter L\'evai$^{1,2}$ and \ Ulrich Heinz$^2$ \\ \\
 ${}^{1}$KFKI Research Institute for Particle and Nuclear Physics, \\
 \ \ P. O. Box 49, Budapest, 1525, Hungary \\
 ${}^{2}$Institut f\"ur Theoretische Physik, Universit\"at Regensburg, \\
 \ \ D-93040 Regensburg, Germany }

\date{24 October 1997}

\maketitle


\begin{abstract}
We analyze recent results of SU(3) lattice QCD calculations with a 
phenomenological parametrization for the quark-gluon plasma equation of 
state based on a quasi-particle picture with  massive quarks and gluons.
At high temperature we obtain a good fit to the lattice data using 
perturbative thermal quark and gluon masses from an improved HTL scheme.
At temperatures close to the confinement phase transition the fitted masses
increase above the perturbative value, and a non-zero (but small)
bag constant is required to fit the lattice data.
\end{abstract}

\section{INTRODUCTION}
\label{sec1}

Strong interactions are described by SU(3) Yang-Mills field theory,
and the fundamental degrees of freedom of Quantum Chromodynamics (QCD)
are gluons and quarks. At high temperature this theory is weakly 
coupled, allowing for the use of perturbative methods. The leading order 
contributions to the characteristic collective excitations and the
equation of state of a quark-gluon plasma were already determined many
years ago \cite{Kapusta,Klim,Weldon}. A gauge invariant approach to
non-leading corrections was derived more recently in form of the
Hard Thermal Loop (HTL) approximation \cite{HTL1,HTL2,HTLmass}
and its improved versions \cite{Rebh,impHTL}.

The theory of QCD predicts the appearance of a phase transition
between the quark-gluon dominated high energy region and the hadronic
state in the low energy region. The two states are characterized by a
dramatic difference in the number of degrees of freedom. Perturbative
QCD can describe successfully only the asymptotic state at very high
momenta, but it fails close to this phase transition where
non-perturbative effects become dominant. Where the perturbative
region begins is still a matter of debate \cite{Kaj97}. The numerical
method of lattice QCD can describe both sides of this phase
transition. Recent developments of this field have yielded dramatic  
improvements both in the extrapolation to the continuum limit and for 
the inclusion of dynamical fermions. 

The existence of this phase transition generated a large experimental
effort to create it in the laboratory and investigate it in detail
through heavy ion collisions. However, to verify the appearance of a
deconfined state in such experiments we need accurate knowledge about
its subsequent hadronization. There is a general interest to
create phenomenological models for this phase transition which agree
with the perturbative results at high energy and with the lattice QCD
data at low energies. Such models must specify the basic degrees of
freedom in the plasma state which will participate in the formation of
hadrons during the confinement phase transition. The experience from
phase transitions in solid state physics and other fields suggests the
introduction of quasi-particles with effective masses generated
through the interactions among the basic constituents. If a large part
of the interaction can be included into the effective masses, then a
quasi-particle description of the interacting matter can be generated
in which the quasi-particles move freely or interact only weakly
\cite{Gorens}. Such a model would be a preferable starting point for
phenomenological hadronization models and for a description of the
strongly interacting matter near the phase transition.

Our goal is to identify the appropriate degrees of freedom through an
analysis of lattice QCD data. Similar attempts started as soon as good
lattice QCD results on the equation of state of strongly interacting
matter became available, and they developed in parallel with the
improvements of the lattice data. It was observed already quite some
time ago that the formula for the free energy density of a massive
ideal gas gives a quite satisfactory description of the numerical
lattice QCD simulations \cite{Ka89}. In Ref.~\cite{KaMeSa88}
temperature dependent screening masses and coupling constants were
extracted from SU(2) lattice data \cite{EngSU289} and compared with
perturbative results. In Ref.~\cite{GoSa93} it was shown that the
obtained thermal mass can be indeed used as an effective mass in the  
equation of state, comparing favorably with SU(2) lattice QCD data.
Early (very poor) lattice data on the pressure and energy density of
pure SU(3) gauge theory in the temperature region $1.2 < T/T_c < 2.4$
\cite{UkaSU389} were interpreted with a constant gluon mass, 
$M_g \approx 500$ MeV, and a constant bag constant 
\mbox{$B^{1/4} \approx 200$ MeV} \cite{BLM90}. Newer SU(3) lattice
data of better quality \cite{EngSU390} made a reanalysis of the
gluonic equation of state possible which yielded a new expression for
the temperature dependent coupling constant in the thermal gluon mass
\cite{Pesh94}. A quasi-particle based, thermodynamically consistent
analysis of the SU(2) \cite{EngSU289} and SU(3) \cite{EngSU390}
lattice data was performed in Ref.~\cite{Gorens} whose authors
investigated in detail the temperature dependence of the thermal gluon
mass and of the bag constant. New SU(3) lattice data with a complete
continuum extrapolation appeared in Refs.~\cite{Boyd95,LaerQM96},
and again a phenomenological analysis proved the applicability of an
equation of state with massive gluons \cite{Pesh96}. Furthermore it
was shown \cite{Pesh96} that at $T> 2.5 \ T_c$ the Debye screening
mass extracted from lattice QCD correlation functions and the thermal
gluon mass fitted to the lattice QCD equation of state are consistent
with each other, such that at high temperature the perturbative QCD
with effective massive degrees of freedom provides a good description
of the lattice QCD results \cite{Pesh96}. In Ref.~\cite{Pesh96} the
lattice data fit yielded  $d=17.2$ gluon degrees of freedom,
supporting the presence of massive {\em transverse} modes only. A
direct investigation of the screening mass in SU(3) lattice
calculations was performed in Ref.~\cite{Gros94} on both sides of the
phase transition. 

A somewhat different analysis of pure SU(3) lattice data was presented
in Ref.~\cite{EngSU390}. It assumes the existence of massless gluons 
above a certain minimal momentum (``cut-off model'') together with
glueball-like non-perturbative massive excitations. This model could
reproduce the SU(3) lattice data quite well \cite{EngSU390,Ris92}.

All of the above analyses were performed for the case of pure SU(2)
and SU(3) gauge theories, because sufficiently high-quality lattice
QCD data existed only in these cases (for SU(3) the latest results can
be found in Refs.~\cite{Boyd95,LaerQM96}). However, new lattice data
including dynamical fermions appeared recently, with $N_f=2$
\cite{SU3NF2a,SU3NF2b} and $N_f=4$ \cite{SU3NF4}  quark
flavors. Extended investigations are under evaluation for the cases
$N_f=1,3,6$ and $N_f \geq 7$ (see Ref.~\cite{LatQCD}), but in their
present state they can not yet be used to extract temperature
dependent thermal masses. These new lattice results can now be used
for the extraction of a phenomenological equation of state for the
quark-gluon plasma (QGP) containing massive gluons and quarks, which
to our knowledge has not been published elsewhere. Similar work is
under way in the group of A. Peshier and B. K\"ampfer \cite{Pesh97}.

The understanding of the dynamical generation of effective quark and
gluon masses could be very important in many research fields. 
Dynamical mass generation is an essential ingredient for the 
solution of the infrared catastrophe in hot gauge theories
\cite{Braat95} and the formulation of effective field theoretical
approaches to thermal QCD \cite{BraNi95}. It is at the heart of recent 
new approximation schemes such as the ``screened perturbation theory''
\cite{KaPaPe,foamDrum}. Screening masses have also successfully been
used to regulate the low-$p_T$ behavior in the parton cascade approach
to ultrarelativistic heavy ion collisions \cite{MuWa}. Massive quarks
are the basic degrees of freedom in hadronization models: the
phenomenological model ALCOR is based on the coalescence of massive
quarks and antiquarks into hadrons \cite{ALCOR}; transport
descriptions of the hadronization based on the Nambu-Jona-Lasinio
model contain massive quarks as well \cite{Klev}. 
Massive quarks and gluons obtained from HTL approximation \cite{HTLmass}
were already applied to estimate charm production at RHIC and LHC energies
\cite{LevVo}.

In Sec.~\ref{sec2} we summarize the present knowledge about the
thermal and screening masses in perturbative thermal QCD. In
Sec.~\ref{sec3} we overview the equation of state of quasi-particles,
its most important thermodynamical properties and discuss the
criterium of thermodynamical consistency. In Sec.~\ref{sec4} we
analyze pure SU(3) lattice QCD results; we determine the relevant
number of gluonic degrees of freedom, the temperature dependent gluon
mass and coupling constant, the temperature dependent bag constant and
investigate the screening mass. In Sec.~\ref{sec5} we repeat this
analysis on lattice data with $N_f=2$ and $N_f=4$ dynamical fermion
species. In Sec.~\ref{sec6} we generate the equation of state of the
QGP in general and investigate the speed of sound in this system. In
Sec.~\ref{sec7} we include hadronic matter into our investigation and
describe phenomenologically the phase transition between QGP and
hadronic matter on the basis of our new equation of state obtained
from lattice QCD data. In Sec.~\ref{sec8} we discuss our results.

\section{THERMAL AND SCREENING MASSES IN \\ 
         PERTURBATIVE QCD}
\label{sec2}
\subsection{Gluons}
\label{sec2.1}

There are essentially two ways to define an effective dynamical
gluon mass: via the pole of the effective gluon propagator,
or via the long-range behavior of the potential between two 
heavy color sources. In both cases one has to address the question of
gauge invariance of the result for the mass, since the defining
objects are not themselves gauge invariant.

On a perturbative level, it was noticed by Klimov \cite{Klim}
and Weldon \cite{Weldon} that the leading term in a high-temperature
expansion of the 1-loop gluon polarization tensor was gauge invariant.
It was later shown by Heinz \cite{Uli86} that the same result for the
gluon polarization tensor could be obtained from classical color
kinetic theory in the linear response approximation. From this
expression for the polarization tensor Weldon \cite{Weldon} derived
the following dispersion relation for transverse gluons with momenta 
$\omega, k \gg gT$:
\begin{eqnarray}
\omega^2 &=& k^2 + M_{g,\infty}^2(T) \ , \label{weld1} \\
M_{g,\infty}^2(T) &=& \frac{g^2 T^2}{2} \left( \frac{N_c}{3}
+ \frac{N_f}{6} \right)  \ . \label{infmg}
\end{eqnarray}
One might object that the high temperature limit in which the 
polarization tensor was obtained is inconsistent with the limit
$\omega, k \gg gT$ under which Eq. (\ref{weld1})
was derived. However, Rebhan et al. showed recently 
\cite{Rebh,impHTL} that the validity of Eq. (\ref{weld1}) does not
depend on the high temperature limit taken in Ref.~\cite{Weldon};
indeed, near the light cone the transverse 1-loop polarization
function is given by the gauge invariant result
 \begin{equation}
  \Pi_t(\omega^2=k^2) =  M_{g,\infty}^2(T)
 \end{equation}
independent of the magnitude of $\omega=k$ relative to $T$. Since for
large momenta $k \gg gT$ the gluon dispersion relation moves
arbitrarily close to the light cone, $M_g(T)$ can be interpreted as
the thermal gluon mass in the high momentum limit. Its inclusion as a
gluon mass term in higher orders of perturbation theory removes
certain classes of collinear singularities near the light cone, and
within the context of Hard Thermal Loop resummation
\cite{HTL1,HTL2,HTLmass} it leads to the so-called ``improved HTL
resummation scheme'' \cite{impHTL}.

The ``Debye screening mass'' $M_D$, on the other hand, is related to
the behavior of gluonic excitations at small momenta. It can be
defined either through the static limit of the gluon polarization
tensor $\Pi_{\mu \nu} (\omega=0,k)$ for $k\rightarrow 0$ or via the
behavior of the potential between two heavy color charges at large
distances, $V(r) \sim \exp(-M_D r)/r$ \cite{HTLmass}. In the first
case a gauge invariant result can be obtained from the leading  
term of the high temperature expansion for the 1-loop gluon polarization
operator \cite{Klim,Weldon} or, equivalently, from the 1-loop result
within the HTL resummation scheme \cite{HTL1,HTL2,HTLmass} for the
gluon polarization operator at momenta $\omega, k < gT$. In the second
case, care must be taken to make the external sources gauge invariant 
\cite{HTLmass}.

In the HTL approximation one obtains from the behavior of \\
\mbox{$\Pi_L(\omega=0,k\rightarrow 0)$} the Debye mass
\cite{HTL1,HTL2,RebDeb} 
 \begin{equation}
   M_D^2 = 2  M_{g,\infty}^2(T) = g^2 T^2 \left(\frac{N_c}{3} +  
   \frac{N_f}{6} \right) \ .
 \end{equation}
This result is related to the 1-loop plasmon frequency, $\omega_p$,
obtained from the gluon dispersion relation (pole of the propagator)
for $k \rightarrow 0$, by $M_D^2 = 3\, \omega_p^2$.

\subsection{Quarks}
\label{sec2.2}

Similarly to the gluon propagator, one can study the quark propagator
in the 1-loop approximation. In the high temperature or low-momentum
limit $\omega, k \ll T$ \cite{Klim,Weldon} or, equivalently, in the
HTL approximation \cite{HTL1,HTL2} one obtains
 \begin{equation}
   M^2_f(T) = \frac{N_c^2 -1}{2N_c} \frac{g^2 T^2}{8} =
   \frac{g^2 T^2}{6} \, .
 \label{HTLmq}
 \end{equation}
This can be interpreted as the effective quark mass for soft quarks with
momenta $\omega,k < gT$. For high momenta $\omega,k \gg gT$ the fermion
dispersion relation again approaches the light cone, and one can make
use of the gauge invariant light cone limit of the 1-loop fermion
self energy \cite{impHTL}:
\begin{eqnarray}
\Sigma_f(\omega^2=k^2) &=&  M^2_{f,\infty}(T) \ , \\ 
M^2_{f,\infty}(T) &=& 2 M_f^2(T) =
\frac{g^2 T^2}{3} \ . \label{infmq}
\end{eqnarray}
Again this is an exact 1-loop result, independent of the value of
$\omega=k$, and $M_{f,\infty}$ can be interpreted as the thermal
quark mass for high momentum quarks, $k \gg gT$.

\subsection{Temperature dependent coupling constant}
\label{sec2.3}

The thermal gluon and quark masses obtained from perturbative QCD
are displayed in Eqs.~(\ref{infmg}) and (\ref{infmq}). All masses
depend on the temperature $T$ and the strong coupling constant $g$. 
At higher orders of the loop expansion, the coupling $g$ begins to run
as a function of $T$, giving rise to a temperature dependent
effective coupling constant $g(T)$. In $SU(N_c)$ gauge theory at
$T=0$, in the presence of $N_f$ quark flavors the 1-loop expression
for the running coupling constant as a function of the momentum
transfer $Q$ is 
 \begin{equation}
   g^2(Q^2) = \frac{48 \pi^2}{(11 N_c - 2 N_f) \ln (-Q^2/\Lambda_N^2)}
   \, .
 \label{gfun0}
 \end{equation}
Here $\Lambda_N$ is the cut-off parameter. It was shown in
Ref.~\cite{KajKap85} that in a thermal system it makes sense to
introduce a temperature dependent function $g^2(T)$ by the following
parametrization: 
 \begin{equation}
   g^2(T) = \frac{48 \pi^2}{(11 N_c - 2 N_f) \ln F^2(T,T_c,\Lambda)} 
   \, .
 \label{gfun1}
 \end{equation}
We expect $F(T,T_c,\Lambda)$ to be linear in the region where
perturbation theory is valid \cite{KajKap85}. In other words, writing
 \begin{equation}
   F(T,T_c,\Lambda) = K(T/T_c) \cdot \frac{T}{T_c} \cdot
   \frac{T_c}{\Lambda_{\overline {MS}}}
 \label{gfun2}
 \end{equation}
we expect $K(T/T_c)$ to be a constant in the perturbative region,
i.e. at high $T$. At low $T$ a possible $T$-dependence of $K$
reflects non-perturbative corrections.

In SU(3) at different $N_f$ we have different critical temperatures $T_c$
and different normalization parameters 
${T_c}/{\Lambda_{\overline {MS}}}$ \cite{FinHelKa}:
\begin{eqnarray}
{\rm SU(3), \ } N_f=0 \ \longrightarrow & T_c = 260 \ MeV ;& \ \ \  
   \frac{T_c}{\Lambda_{\overline {MS}}}=1.03 \pm 0.19 \ ; \label{tla1} \\
{\rm SU(3),\ } N_f=4 \ \longrightarrow & T_c = 170 \ MeV ;& \ \ \  
   \frac{T_c}{\Lambda_{\overline {MS}}}=1.05  \ . \label{tla2} 
\end{eqnarray}
For $N_f=2$ the critical temperature in SU(3) was determined as
$T_c= \ 140$ MeV \cite{SU3NF2a,SU3NF2b}, but the normalization factor
${T_c}/{\Lambda_{\overline {MS}}}$ is not known by the authors. From
Eqs.~(\ref{tla1}) and (\ref{tla2}) we will estimate this value
by interpolation and we will use ${T_c}/{\Lambda_{\overline {MS}}}=1.03$.

We will try to fit the lattice data with a non-interacting gas
of massive quarks and gluons, with masses given by the perturbative
expressions (\ref{infmg}) and (\ref{infmq}), but with $g^2$ replaced
by a phenomenological running coupling $g^2(T)$. Using
Eqs.~(\ref{gfun1}) and (\ref{gfun2}) we can then extract a function
$K(T/T_c)$. In a very early work \cite{KaMeSa88} in pure SU(3) gauge
theory this function $K(T/T_c)$ was determined from a similar fit to
the heavy quark-antiquark potential; the fit result was a constant,
${\hat K}(T/T_c) = 19.0$. However, the authors of Ref.~\cite{KaMeSa88}
used $T_c/\Lambda_{MS}=1.78 \pm 0.03$ which differs strongly from
the now accepted value. With our normalization (\ref{tla1}),
${\hat K}(T/T_c) = 19.0$ would correspond to $K(T/T_c) = 33.8$.

Another result was obtained recently from a numerical fit of the
equation of state in pure SU(3) gauge theory \cite{Pesh96}: 
\begin{equation}
F(T,T_c,\Lambda) = 4.17 \frac{T}{T_c} -2.96 \ .
\end{equation}
Obviously, this parametrization differs from Eq.~(\ref{gfun2}).

Our expectation is to obtain a constant $K(T/T_c)$ function in that
temperature region where perturbative QCD and the quasi-particle
picture of the lattice QCD results overlap. A strong temperature
dependence of $K(T/T_c)$ implies non-perturbative effects. We
summarize our results on the function $K(T/T_c)$ for different values
of $N_f$ in Sects.~\ref{sec4} and \ref{sec5}. 

\section{EQUATION OF STATE  WITH QUASI-PARTICLES}
\label{sec3}

In this Section we introduce the equation of state of a pure gluon
plasma and a quark-gluon plasma consisting of massive
quasi-particles. This equation of state (resp. its parameters) will be
determined from the SU(3) lattice data. The key point is to consider
temperature dependent effective masses, $M_i(T)$, which are
dynamically generated. We can introduce the dispersion relation for
quasi-particles: 
\begin{equation}
\omega_i^2(k) =k^2+M_i^2(T)\ . \label{disp}
\end{equation}
Here $k$ is the momentum of the quasi-particle 
and $\omega_i(k)$ is its energy.
A pure gluon plasma  contains only gluons ($i=g$, $N_f=0$), 
a QGP consists of quarks and antiquarks also
($i=g,q,{\overline q}$, $N_f > 0$).
We can define  the Bose  or
Fermi distribution functions $f_i(k)$ for the quasi-particles as
\begin{equation}
f_i(k) = [\exp\{\sqrt{k^2+M_i^2(T)} /T\} \pm 1 ]^{-1}
= [\exp (\omega_i /T ) \pm 1 ]^{-1} \ . \label{dist}
\end{equation}

In addition to the effective masses $M_i(T)$ we also need
the effective number of degrees of freedom of the quasi-particles
to determine pressure and energy density. The correct value for the
number of effective 
gluonic degrees of freedom is not obvious, since a
massless vector boson has two helicity states, but
massive vector bosons have three spin states.
One possibility is to introduce a temperature dependent
effective degeneracy factor $D(T)$ for the gluons. We will
determine the value of $D(T)$ in a pure gluon plasma ($N_f=0$),
where quark degrees of freedom do not interfere. Later we can
assume the same value for $D(T)$ in a QGP.

Another possibility is to fix the number of gluonic degrees of freedom 
(e.g. at its perturbatively expected value $D_g=16$) as well as the
number of quark and antiquark degrees of freedom. In this case we have
the freedom to introduce an effective interaction term for those
contributions of the strong interaction which can not be absorbed by
the presence of effective masses. We parametrize this term by a
temperature dependent ``bag constant'', $B(T)$. This interaction term
$B(T)$ can be determined in a pure gluon plasma as well as in a
quark-gluon plasma. As we will see in the following Section, in the
QGP phase quarks and anti-quarks will also contribute to $B(T)$, so
its physical meaning differs from the bag constant parameter of the
MIT Bag Model; nevertheless we will use the same notation for
convenience. 

\subsection{The number of effective gluonic degrees of freedom $D(T)$}
\label{sec3.1}

Let us first consider pure SU(3) gauge theory where we wish to find
out the number of contributing gluonic quasi-particle degrees of
freedom, $D(T)$. A strongly interacting gluon plasma contains not only
the transverse gluon modes, but at low momenta $k<gT$ there exist also
longitudinal plasma excitations. It was shown in Ref.~\cite{HTL1} that
at high momenta $k \gg gT$ the longitudinal modes disappear, in the
sense that the residue of the corresponding pole in the propagator
becomes exponentially small. Since the equation of state is dominated
by particles with momenta $k \sim T$, we expect at high temperatures 
$T\rightarrow \infty$ (where $g$ becomes small) the contribution of
the longitudinal modes to be negligible. On the other hand, at low
temperatures (where $g$ can become of order 1 or larger) their
contribution may be relatively large. Therefore we can not a priori
neglect them from the analysis of lattice QCD data. 

We will introduce a temperature dependent effective number of gluon
degrees of freedom $D(T)$, generalizing the approach of
Ref.~\cite{Pesh94} where $D$ was taken as a constant (temperature
independent) fit parameter. We expect that in the high temperature
limit there are only transverse gluonic quasi-particle  modes,
yielding $D(T=\infty)=D_g=16$. At low temperatures the longitudinal
gluons are expected to also contribute, giving $D(T)> D_g$.

With these assumptions the thermodynamical pressure $P_g(T)$ of pure
gluon matter (which is nothing but the grand canonical thermodynamical
potential) can be written as 
 \begin{equation}
   P_g(T)= \frac{D(T)}{(2\pi)^3} \int_0^\infty d^3k 
   \frac{k^2}{3 \omega_g(k)} f_g(k) \ . \label{dpt}
 \end{equation}
{} From the thermodynamic identity 
 \begin{equation}
   \varepsilon(T)=T \cdot  \frac{d P(T)}{d T} -P(T) \label{entr}
 \end{equation}
(we work at zero net baryon density, $\mu_B=0$)
one obtains the following relation for the energy density:
 \begin{eqnarray}
   \varepsilon_g(T)&=&\frac{D(T)}{(2\pi)^3} \int_0^\infty d^3k \ 
   \omega_g(k) \ f_g(k) 
     + W_D(T) \ , \\
   W_D(T)&=& T \frac{P_g(T)}{D(T)} \frac{d  D(T)}{d T} 
   - T M_g(T) \frac{d   M_g(T)}{d  T}\ 
   \frac{D(T)}{(2\pi)^3} \int_0^\infty \frac{d^3k}{\omega_g(k)} \ f_g(k) 
    \ . \ \hspace*{0.3 truecm}
 \label{det}
 \end{eqnarray}
Here $W_D(T)$ summarizes the extra terms stemming from the temperature 
derivative of the effective mass and effective number of degrees of
freedom. On the other hand, a consistent quasi-particle picture
demands $W_D\equiv 0$ \cite{Gorens}. Requiring this identity yields a
differential equation for $D(T)$: 
 \begin{equation}
   D(T)=D(T^*) + 
   \int_{T^*}^T dT' M_g(T') \frac{dM_g(T')}{dT'} \ 3 D(T')
   { {\int_0^\infty \frac{d^3k}{\omega_g(k)} f_g(k) } \over
     {\int_0^\infty \frac{d^3k ~k^2}{\omega_g(k)} f_g(k) } } \ .
 \label{dt} 
 \end{equation}
Here $T^*$ is an integration constant. If the functions $M_g(T)$ and
$D(T)$ satisfy this selfconsistency condition, then we have a
thermodynamically consistent quasi-particle description. 

Lattice calculations yield numerical values for the functions $P_g(T)$
and $\varepsilon_g(T)$. These two sets of data are sufficient to
determine for each value of $T$ the quantities $M_g(T)$ and $D(T)$
such that they satisfy Eqs.~(\ref{dpt}-\ref{det}) with $W_D=0$. Since
the lattice results were created in a thermodynamically consistent way
we expect that the resulting functions  $M_g(T)$ and $D(T)$ will
satisfy Eq.~(\ref{dt}). We will check this as a test for the
consistency of our extraction procedure for $M_g(T)$ and $D(T)$. Our
numerical results on $M_g(T)$ and $D(T)$ will be summarized in
Sec.~\ref{sec4}. 

\subsection{The temperature dependent interaction term $B(T)$}
\label{sec3.2}

We can also investigate the high temperature limit from another point
of view \cite{Gorens,Pesh96}. Let us fix the value of the gluonic
degrees of freedom at every temperature to $D(T)\equiv D_g = 16$,
assuming that only transverse gluons contribute to the thermodynamic
quantities, and calculate directly any extra contribution to the
pressure and the energy density. In this way both pure gluon matter
and a quark-gluon plasma can be investigated. We will here consider a
QGP with $N_f$ quark flavors and the usual number of quark degrees of
freedom: $D_q=D_{\overline q}=2 \cdot 3 \cdot N_f = 6 N_f$.

The effective quark mass $M_q(T)$ will be related to the effective
gluon mass $M_g(T)$ through the value of the temperature dependent
coupling constant $g(T)$ via the perturbative expressions in
Eqs.~(\ref{infmg}) and (\ref{infmq}) for the effective masses. This
means that we absorb non-perturbative features into a common
non-perturbative fit function $g(T)$, without touching the
perturbative form of the Eqs.~(\ref{infmg}) and (\ref{infmq}) itself. 

The deviation of $\varepsilon_0(T) = \sum_i \varepsilon_i(T)$ 
and $P_0(T)=\sum_i P_i(T)$ from the
ideal gas values corresponding to the effective masses $M_i(T)$
and the fixed degeneracy factors $D_g$, $D_q$ and $D_{\overline q}$
will be parametrized by a temperature dependent function $B(T)$.
At high temperature, where we expect a perturbative picture based on 
free quarks and transverse gluons to be valid, $B(T)$ should
vanish. At lower temperature, $B(T)$ provides another measure for
non-perturbative physics which can not be absorbed into effective 
quark and gluon masses. Both quarks and gluons contribute to $B(T)$.

We introduce $B(T)$ by writing the pressure in the form
 \begin{equation}
   P(T)= \sum_{i=g,q,{\overline q}} 
   \frac{D_i}{(2\pi)^3} \int_0^\infty d^3k 
   \frac{k^2}{3 \omega_i(k)} f_i(k) - B(T) \ , \label{pt}
 \end{equation}
motivated by the MIT Bag Model. Thermodynamic identities yield the
following relation for the energy density: 
 \begin{equation}
   \varepsilon(T)= \sum_{i=g,q,{\overline q}}
   \frac{d_i}{(2\pi)^3} \int_0^\infty d^3k \ \omega_i(k) \ f_i(k)
   + B(T) + {\widetilde W}_B(T) \ . \label{et}
 \end{equation}
Here ${\widetilde W}_B(T)$ summarizes the extra terms obtained from 
the temperature de\-ri\-va\-tive of $B(T)$ and of the effective mass
$M_i(T)$. Similarly as above thermodynamic consistency of our 
quasi-particle model requires ${\widetilde W}_B(T) \equiv 0$.
This yields to the following integral representation  for $B(T)$
\cite{Gorens,Pesh96}:
 \begin{equation}
   B(T)= B(T^*) - \sum_{i=g,q,{\overline q}} 
   \frac{ D_i}{(2\pi)^3} \int_{T^*}^T dT' M_i(T') \frac{dM_i(T')}{dT'}
   \int_0^\infty \frac{d^3k}{\omega_i} f_i(k) \ . \label{bt}
 \end{equation}
Again $T^*$ is an integration constant. If this expression is satisfied
by the functions $M_i(T)$ and $B(T)$, we have a thermodynamically
correct quasi-particle description containing effective gluons and
quarks with effective thermal masses. 

Similarly as above, from the numerical values for the functions $P(T)$
and $\varepsilon(T)$ obtained from lattice QCD and assuming
${\widetilde W}_B(T)=0$,  we can determine the appropriate values for
$M_i(T)$ (resp. $g(T)$) and $B(T)$ through Eqs.~(\ref{pt}) and
(\ref{et}). As before we expect  that the obtained functions $M_i(T)$
and $B(T)$ will satisfy Eq.~(\ref{bt}). Once again this is a good
numerical consistency check for our extraction of the functions
$M_i(T)$, $B(T)$. 

\section{ANALYSIS OF LATTICE DATA FOR PURE \\ SU(3) GAUGE THEORY}
\label{sec4}

We analyze the latest lattice data in the continuum limit for pure
SU(3) gauge theory \cite{Boyd95,LaerQM96}. The calculations of 
Ref.~\cite{Boyd95} show a $\pm 2 \%$ error on the data for $P(T)$,
$\varepsilon(T)$ and $s(T)$. We include this error into our analysis;
it will be indicated in Figs.~1, 2 and 3 by dotted lines above and below
the solid lines corresponding to the results extracted from the
mean values.

We begin by determining the effective number of gluonic quasi-particle
degrees of freedom in a gluon plasma, extracting the functions
$M_g(T)$ and $D(T)$ as explained in Sec.~\ref{sec3.1}. Fig.~1a shows
the temperature dependent effective gluon mass, $M_g(T)$, while in
Fig.~1b we display the effective number of gluonic degrees of freedom,
$D(T)$. At high temperature, $T > 2.3 \ T_c$, the number of degrees of
freedom is rather constant, $D(T)=14.5 \pm 1$. Please note that the $\pm
2 \%$ error of the lattice data results in an uncertainty of about 1
gluon degree of freedom. Within this uncertainty the result is
surprisingly close to the naive expectation that at high temperature
only the transverse gluonic modes are present. In the low temperature
region, $T < 2.3 \  T_c$, the function $D(T)$ increases, which one may
wish to associate with the increasing contribution of the longitudinal
modes. However, for $T < 1.5 \ T_c$, $D(T)$ rises very strongly and 
exceeds even the value of $24$ expected for massive gluons with $8$
color and $3$ helicity states. This indicates the presence of strong
non-perturbative effects at low temperature not all of which can be
absorbed by the effective gluon mass. In fact, part of the strong rise
in $D(T)$ is driven by the strong increase of the effective gluon mass
which leads to exponential suppression of the thermodynamical
contributions to $P(T)$ and $\varepsilon(T)$. We consider this
behavior of $D(T)$ as physically unreasonable and believe that a
parametrization of these non-perturbative effects via an extra 
interaction term makes more sense.

To this end we fix the effective number of gluonic modes to $D_g=16$
as suggested by the behavior of $D(T)$ in Fig.~1b and our perturbative
prejudice at large $T$. This means that we consider only transverse
effective gluons as quasi-particle states and subsume all further
interaction effects into a non-perturbative ``bag constant''
$B(T)$. The procedure is described in Sec.~\ref{sec3.2}. Fig.~2a shows
the temperature dependent transverse gluon mass $M_g(T)$  resulting from
this approach. Comparing with Fig.~1a one sees that at large values of
$T$ the effective mass is now somewhat larger, compensating for the choice
of a fixed $D_g=16$ instead of the best fit value $D_g \approx 14.5$ 
in Fig.~1b. For smaller temperatures, however, the rise of $M_g(T)$
is much weaker than in Fig.~1a, and $M_g(T)$ also shows a much weaker
sensitivity to the statistical error of the lattice data on $P(T)$ and 
$\varepsilon(T)$. This indicates that this type of parametrization
provides a more reasonable fit to the lattice data than the one above
in terms of $M_g(T)$ and $D(T)$. Please note that the effective gluon
masses shown in Fig.~2a correspond in the region $T>2 \ T_c$ to a
coupling constant $g=\sqrt{2} M_g(T)/T \approx 1.3-1.4$, in agreement
with other extraction procedures: for example, the authors of
Ref.~\cite{Interq} obtained from the interquark potential at small
distances the value $\alpha \approx 0.15$ which translates into 
$g=1.37$.

Fig.~2b shows the function $B(T)/\varepsilon(T)$, the interaction term
relative to the total energy density. At high temperatures, $T > 2.3\
T_c$, the bag constant is very small, only about $2-3 \%$ of the total
energy density, although slightly negative. We take the smallness of
$B(T)$ as confirmation for the validity of a quasi-particle picture at
large temperature. Near $T= 2.3 \ T_c$ $B(T)$ changes sign, increasing
for smaller values of $T$, but never exceeding a value of $15 \%$ of
the total energy density even at $T_c$. Note that this relatively
small deviation from an ideal quasi-particle picture gave rise to 
the dramatic and unphysical behavior of $D(T)$ (and $M_g(T)$) in the
alternative approach above. Clearly, the parametrization via $B(T)$ is
more economical and leads to a more selfconsistent picture.

The dashed line in Fig.~2b shows the result of an integration of
Eq.~(\ref{bt}), using the extracted $M_g(T)$ from Fig.~2a and choosing
boundary conditions $B(T^*)=0$ at $T^*=2.3  \ T_c$. The agreement with
the numerically extracted values for $B(T)$ is nearly perfect
indicating the thermodynamical consistency of the lattice data and our
extraction procedure. 

{} From the temperature dependent gluon mass, $M_g(T)$, one can determine
the temperature dependent coupling constant, $g(T)$, and extract the
function $K(T/T_c)$ given in Eq.~(\ref{gfun2}). The result for pure
SU(3) gauge theory is shown in Fig.~3. The dotted lines again indicate
the $\pm 2 \%$ systematic error of the lattice data. One can see that
at high temperature we obtain a constant value, $K(T/T_c \rightarrow
\infty)= 18 \pm 8$. At low temperature $K(T/T_c)$ decreases indicating 
a larger coupling constant and a larger thermal gluon mass. Close to
the phase transition  the function approaches the value $K(T/T_c
\rightarrow 1)= 1.5$; the precise value depends, however, on the
latent heat at the phase transition point which is still under
intensive investigation in lattice QCD calculations.

The obtained curve can be fitted rather well with the following function
(dashed line in Fig.~3):
 \begin{equation}
   K(T/T_c) = \frac{18}{18.4 \cdot e^{-0.5  (T/T_c)^2} + 1 } \, .
 \label{kfun}
 \end{equation}
If we start from this approximate function of $K(T/T_c)$ for the
coupling constant $g(T)$, we obtain an analytically parametrized,
approximate thermal gluon mass $M_g(T)$ (see
Eqs.~(\ref{infmg},\ref{gfun1},\ref{gfun2})). Once we fix the boundary
condition $B(T^*)=0$ at $T^*=2.3 \ T_c$ as determined from the
numerical analysis of the lattice data (see Fig.~2b), the bag constant
$B(T)$ can be uniquely reconstructed by integrating Eq.~(\ref{bt}).
With $M_g(T)$ and $B(T)$ known, $P(T)$ and $\varepsilon(T)$ are easily
evaluated using Eqs.~(\ref{pt},\ref{et}) (with ${\widetilde W}_B=0$).
This means that Eq.~(\ref{kfun}) together with the value for
  $T^*$ provide a complete analytical parametrization of the equation
  of state.  In Fig.~4 we compare the original lattice QCD data
(diamonds) with the  thus reconstructed values (dashed and
dash-dotted lines). The dash-dotted lines neglect the interaction term
$B(T)$ and thus represent only the quasi-particle contribution to $P$
and $e$. Clearly the agreement between the model and the lattice data
is at most qualitative in this case. Including the bag term, however,
(dashed lines) we obtain nearly perfect reproduction of the lattice
data. Fig.~4 demonstrates the usefulness of a thermodynamically
consistent quasi-particle picture. The deviation from the
Stefan-Boltzmann values indicates that the massless degrees of freedom
can not provide a satisfactory description, while the quasi-particle
model plus bag term leads to qualitative improvements.

\section{ANALYSIS OF SU(3) LATTICE DATA FOR \\ $N_f=2,4$}
\label{sec5}

Newest lattice calculations including dynamical fermions can be used
to determine the equation of state for a realistic quark-gluon plasma
including light quarks. The numerical lattice QCD results can be found
in Refs.~\cite{SU3NF2a,SU3NF2b} for $N_f=2$ and in Ref.~\cite{SU3NF4}
for $N_f=4$. Following these articles we interpolated the expected
continuum result in both cases. From Ref.~\cite{SU3NF2a} for $N_f=2$
we used the data for a current quark mass $am_q=0.0125$ (octagons in
Figs.~9 and 10 in Ref.~\cite{SU3NF2a} and circles in Fig.~7 here). The 
statistical error of the lattice data is small, $< 1 \%$ for the
pressure and $\approx 2-4 \%$ for the energy density. From
Ref.~\cite{SU3NF4} for $N_f=4$ we used the data set for a current
quark mass $m_q/T=0.2$ and we considered their extrapolation of the
energy density to the chiral limit (see $\Box$ in Fig.~4 of
Ref.~\cite{SU3NF4} and in Fig.~7 here). In this case the statistical
errors are even smaller,  but the systematic errors resulting from
  the extrapolations to the chiral and continuum limits are hard to
  judge. To estimate the possible influence of errors on the lattice
data for our extraction procedure we will consider a universal $\pm 2
\%$ error in both cases; the corresponding uncertainty will be
indicated by dotted lines in Figs.~5 and 6.

In the framework of perturbation theory the effects of dynamical
quarks on the gluon dynamics are small and can be simply included into
the thermal gluon mass and Debye screening length. No additional
gluonic collective modes arise. However, at small momenta there are
additional fermionic collective modes, the ``plasminos''
\cite{HTLmass}. Based on our experience in pure gluodynamics they are
not expected to contribute to the equation of state at high
temperature; again the residues of their poles in the quark propagator
vanish exponentially for moments $k \gg gT$ \cite{HTL1}. We will
therefore assume a fixed number of degrees of freedom, given by the
perturbative values at high $T$, $D_g=16$, $D_q = D_{\overline q}= 6
N_f$, and include all collective interaction effects into an
interaction term $B(T)$ as for the purely gluonic system above.

The obtained results are qualitatively very similar to the previous
ones for the purely gluonic case. Fig.~5a shows the obtained results
for the effective gluon mass, $M_g/T$. Since $\frac{M_g(T)}{T} =
\frac{g(T)}{\sqrt{2}}\sqrt{1+\frac{N_f}{6}}$, Fig.~5a can be viewed as
giving the temperature dependence of the effective coupling constant
$g(T)$ which according to Eq.~(\ref{infmq}) also determine the
effective quark mass. From Fig. 5 and Eq.~(\ref{infmg}) we extract $g
\approx 1.6$ in the region $T > 2 \ T_c$ for $N_f=4$, somewhat larger
than for the purely gluonic case. 

Fig.~5b shows the interaction term relative to the total energy
density $B(T)/\varepsilon(T)$ both for $N_f=2$ and $N_f=4$. Similarly
to the result in pure SU(3) gauge theory, the interaction term $B(T)$
remains small at $T>2-2.3 \ T_c$, and does not exceed $15 \%$ of the
energy density in the low temperature region. In the case $N_f=2$ the
lattice data allow us to extract $M_g(T)$ and $B(T)$ only in the
temperature region $ T_c < T < 1.4 \ T_c$, however the characteristics
of the obtained curves are close to the results obtained for $N_f=4$.

Fig.~6 displays the obtained functions $K(T/T_c)$ (solid lines)
together with the expected errors (dotted lines) stemming from an
assumed typical $\pm 2 \%$ uncertainty in the lattice data. For all
three cases, $N_f=$ 0, 2 and 4, the functions $K(T/T_c)$ are very
similar (within the admittedly large error bars). Within the existing
uncertainties we can thus define a universal function $K(T/T_c)$;
differences due to variations of $N_c$ and $N_f$ can be largely
absorbed into $T_c$ and $\Lambda_{\overline{MS}}$ in
Eq.~(\ref{gfun2}). It would be very interesting to see whether the
existence of such a universal function $K(T/T_c)$ is confirmed by
future higher quality lattice data. 

Since the data for purely gluonic SU(3) span a much larger range
than those including dynamical fermions, we will use the analytical
form (\ref{kfun}) extracted from the gluonic case also in our
quasi-particle picture for the full QGP. For the zero point of $B(T)$
we will again choose the value $T^*=2.3$ determined in the purely
gluonic SU(3) case, although (modulo large numerical uncertainties)
the $N_f \neq 0$ data seem to favor a somewhat smaller value.

For a QGP with $N_f=2$ quark flavours Fig.~7a shows the lattice data
(circles for pressure and energy density) and the reconstructed values
for $\varepsilon(T)$ and $P(T)$, both including the bag constant
(dashed lines) and neglecting the interaction term (dash-dotted
lines). For $N_f=4$ quark flavours the analogous quantities are
displayed in Fig.~7b. Here the squares give directly the lattice data
from Ref.~\cite{SU3NF4} for the energy density, while the diamonds 
represent the continuous line for the pressure given in Fig.~2 of
Ref.~\cite{SU3NF4}. In each case inclusion of the interaction term
$B(T)$ improves the picture. The agreement is weaker than in the pure
SU(3) case (see Fig.~4), but the pressure is reproduced rather nicely
as well as the energy density at high temperature. Note that here we
used the function $K(T/T_c)$ from Eq.~(\ref{kfun}). The discrepancy in
the region $T \approx T_c$ may be connected to the use of massive
current quarks in the actual lattice simulations and associated
uncertainties in the extrapolation to the chiral limit (for a
discussion see Ref.~\cite{SU3NF4}). In view of those the $10-15 \%$
discrepancy between the reconstructed curves and data is surprisingly
small, especially if we remember that we used here the parametrization
which was obtained from the pure SU(3) case. 

\section{SPEED OF SOUND IN THE MASSIVE QGP }
\label{sec6}

In hydrodynamical models of strongly interacting matter the dynamical
behavior is mostly determined by the value of the speed of sound:
 \begin{equation}
   c_s^2 = \frac{d P}{d \varepsilon} =
   \frac{dP/dT}{d\varepsilon/dT} \ . 
 \label{spe0}
 \end{equation}
Since the energy density $\varepsilon(T)$ was obtained from
  Eq.~(\ref{entr}) we can rewrite this as 
 \begin{equation}
   c_s^2 = {{(P+\varepsilon)/ T}
   \over {d\varepsilon / {dT} } } \ ,
 \label{spe1}
 \end{equation}
  which is more directly connected to the lattice data. For a
  noninteracting ($B(T)=0$) gas of massless particles the speed of
  sound is $c_s^2=1/3$. In Ref.~\cite{LevVo}, using Eq.~(\ref{spe1}), 
  it was found that even for a non-interacting ($B(T)=0$) massive
  quark-gluon gas the speed of sound remains very close to $c_s^2=1/3$. 
  (In that work thermal masses from earlier HTL calculations
  \cite{HTLmass} and a running coupling constant according to
  Eqs.~(\ref{gfun1},\ref{gfun2}) with a constant $K(T/T_c)$ were used.)
If there are, however, strong remaining interactions which cannot be
absorbed in the masses but require a non-vanishing and possibly
strongly $T$-dependent $B(T)$, this will generate deviations of
$c_s^2$ from the ideal gas value, especially near $T_c$
\cite{RischSoft}. A strong drop of $c_s^2$ near $T_c$ will give rise
to a ``soft point'' in the equation of state at which the ability of
the matter to generate expansion flow is minimal. 

Since in our analysis the interaction term  $B(T)$ remains small and
changes smoothly with $T$, we expect only weak  modifications of the
speed of sound around $T_c$. Fig.~8 displays the speed of sound
$c_s^2$ for $N_f=0,2,4$ in our quasi-particle picture and confirms
this expectation. Including the bag term $B(T)$ (dashed lines),
$c_s^2\approx 0.33$ in the high temperature limit, and $c_s^2$
decreases to a minimum value $c_s^2\vert_{min} = 0.15$ around $T_c$.
If we neglect the interaction term $B(T)$ and calculate the speed
  of sound  $c_{s,0}^2$ from the expression
 \begin{equation}
   c_{s,0}^2 = { {dP_0} \over {d\varepsilon_0} } =
   { {d P_0(T) / dT} \over {d\varepsilon_0(T) / dT} } \, ,
 \end{equation}
where $P_0(T)=P(T)+B(T)$ and $\varepsilon_0(T) =
  \varepsilon(T)-B(T)$, then (see dash-dotted line in Fig.~8) the
  speed of sound shows only a very weak temperature dependence,
  decreasing from $c_{s,0}^2\approx 0.30$ at $T = 4.5\, T_c$ to
  $c_{s,0}^2\vert_{min} = 0.23$ at $T_c$. If we instead use
  Eq.~(\ref{spe1}) also in this case (as in Ref.~\cite{LevVo}), where
  $B(T)$ cancels in the numerator, then the resulting speed of sound
  is very close to the full result (dashed line), because of the small
  influence of $B(T)$ on the energy density in the denominator of
  Eq.~(\ref{spe1}). Our result agrees with the calculations in
  Ref.~\cite{LevVo} in the high temperature region and differs around
  $T_c$ because of the non-linear behavior of $K(T/T_c)$ indicating
  non-perturbative effects.

These results on the speed of sound show that a massive quark-gluon
plasma will expand rapidly and that therefore the phase transition
should be a fast process without detectable duration effects. Even
where the pressure already shows large ($>50\%$) deviations from
the massless ideal gas law, the speed of sound still deviates by less
than 10\% from the value $1/\sqrt{3}$. In this way our analysis of
lattice QCD results favors fast hadronization models
\cite{ALCOR,Csor94,Let95} of the deconfined phase. 

\section{GLUON SUPPRESSION AROUND $T_c$ IN SU(3) GAUGE THEORY}
\label{sec7}

In this Section we investigate some phenomenological consequences of
the large effective quark and gluon masses. Fig.~9 displays the
effective masses in GeV for $N_f=0,2,4$. All masses behave
similarly, since they are connected to each other by the function
$K(T/T_c)$ in the effective coupling constant $g(T)$. Just above the
critical temperature the effective masses drop, but then they remain
rather constant in the temperature region $1.2 \ T_c < T < 2.5 \
T_c$. The mass values for the case $N_f=2$ are smaller than for
$N_f=0$ due to the smaller critical temperature $T_c = 140$ MeV
instead of 260 MeV. For $N_f=4$ they increase again due to the larger
contribution from quark loops and the somewhat larger $T_c=170$
MeV. The gluon masses are large in the cases $N_f=0,4$, of order $M_g
= 650-750$ MeV in the temperature region $T_c < T < 3\  T_c$ (solid
lines). In the same temperature region the quark masses are of order
$M_q = 300-400$ MeV (dashed lines). 

Large effective masses yield smaller number density at the same energy
density. Thus the density of these massive quasi-particles will be
smaller than in the massless case. Fig.~10 demonstrates this by
displaying the density ratio of the massive quasi-particles and the
massless ones, $n_i/n_i^0$ (solid lines for gluons and dotted lines
for quarks) as a function of temperature. At large temperature the
ratio flattens and eventually approaches 1, but close to the critical
temperature the ratio drops very quickly to values much smaller than
1. This indicates that the massive quark-gluon plasma is much more
dilute than a massless QGP, especially near $T_c$. 

Now we compare the abundances of quarks and gluons relative to each
other. Eqs.~(\ref{infmg}) and (\ref{infmq}) imply that in our model
the gluon effective mass is always larger than the quark effective
mass by the ratio
 \begin{equation}
   \frac{M_{g, \infty}}{M_{q, \infty}} = \sqrt{\frac{3}{2} 
   \left( \frac{N_c}{3} + \frac{N_f}{6} \right) } \ ,
 \end{equation}
which even increases with the number of quark flavor $N_f$.
Each gluon degree of freedom will thus be suppressed compared to
each quark degree of freedom. After multiplying the corresponding
degeneracy factors we can compute the density ratio for gluons
 \begin{equation}
   R_{g} = \frac{n_g }{n_g +n_q + n_{\overline q}} \ ,
 \end{equation}
which is shown in Fig.~11 as a function of $T$. The horizontal lines
show the ratio $R_{g}^0$ for the massless QGP. The deviation caused by
the effective mass is clearly seen, and a considerable suppression of
the relative gluon number occurs near the phase transition. Near
$T_c$, for $N_f=2$ we have $R_{g}=0.25$ which means that only 25\% of
the particles are gluons while the remaining 75\% are quarks and
antiquarks. For $N_f=4$ the gluon suppression is even stronger and
near $T_c$ we find only 10\% gluons and 90\% quarks and anti-quarks.

These results show that the massive QGP is a dilute system dominated
by massive quarks and  antiquarks. Gluons are much heavier than quarks
and their number densities  are suppressed. These properties indicate
that lattice QCD results support the formation of a quark-antiquark
dominated plasma state just above the critical temperature, which
contains mostly massive quarks and antiquarks. This system hadronizes
quickly. Fast hadronization is also favored by the fact that the
effective quark masses are already close to the masses in the
``constituent quark'' picture of hadrons.  Clusterization of these
massive quarks leads directly to multiquark states with similar masses
as those of the finally observed hadrons, allowing for immediate hadron
formation without having to wait for further energy transfer.   

\newpage

\section{CONCLUSIONS}
\label{sec8}

Lattice data in pure SU(3) gauge theory
and  new lattice results with $N_f=2$ and $N_f=4$ dynamical fermions
suggest that beyond the phase transition temperature strong
non-perturbative effects dominate the deconfined state.
In order to understand this phenomenon and to  facilitate
the application of lattice results in dynamical descriptions
of the evolution of the deconfined phase and its hadronization
we analyzed the lattice data in a phenomenological model
assuming the appearance of massive quasi-particles,
namely massive quarks and massive transverse gluons.
We determined a temperature dependent effective coupling constant
$g(T)$, which determines the effective masses, and the 
interaction term $B(T)$, which summarizes contributions 
not to be included into the effective masses.
We found that such a quasi-particle picture  works very well.
The interaction term $B(T)$ remains small
and the extracted dynamical mass is  
consistent with perturbative QCD results at high temperature
while it includes non-perturbative contributions at low temperature.
In all cases, $N_f=0,2,4$, the obtained characteristics are very 
similar. If we determine $g(T)$ in the purely gluonic case and
extend it in an appropriate way for $N_f>0$, we can reproduce
approximately the $N_f=2,4$ lattice data. This fact suggest
the existence of a universal description of non-perturbative
effects in the fermionless and fermionic cases.
Our analysis leads to a thermodynamically consistent quasi-particle
model for the QGP equation of state which is parametrized
via an effective coupling constant $g(T)$ through 
Eqs.~(\ref{gfun1},\ref{gfun2},\ref{kfun}) and the zero point
$T^*$ of the interaction term $B(T)$.
Further lattice data with improved quality 
over a wider temperature region
are needed to fix $g(T)$ and $T^*$ more accurately.

The successful reproduction of lattice data by this quasi-particle
model indicates the validity of such an approximation. The
consequences of the appearance of these quasi-particles are very
interesting: near $T_c$ lattice results indicate the existence 
of a QCD phase which is dominated by massive quarks and anti-quarks,
while the even heavier gluons are suppressed. This quark-antiquark
plasma can be characterized by a relatively large speed of sound which
indicates fast dynamical evolution and the lack of long time delays
during hadronization. Massive quarks and anti-quarks can form
clusters, 2- and 3-body objects, which can be easily associated 
with the hadronic mass spectra: massive quarks ($M_q=300-350$ MeV)
above the critical temperature turn  into similarly massive
``constituent quarks'' inside hadrons below $T_c$ in a smooth but
rapid hadronization process.

\section*{ACKNOWLEDGMENTS}

Discussions with T.S. Bir\'o, E. Laermann, B. M\"uller, A. Sch\"afer,
M. Thoma, J. Zim\'anyi,  and especially with B. K\"ampfer and
  A. Peshier, are gratefully acknowledged. P.L. is grateful for the
warm hospitality of the Institute f\"ur Theoretische Physik at
Universit\"at Regensburg. This work was supported in part by DAAD
by the National Scientific Research Fund (Hungary)
OTKA Grant Nos. F019689 and T016206 (P.L.), and
by DFG, BMBF and GSI (U.H.).

\newpage
\section*{Figure Caption}

\begin{description}

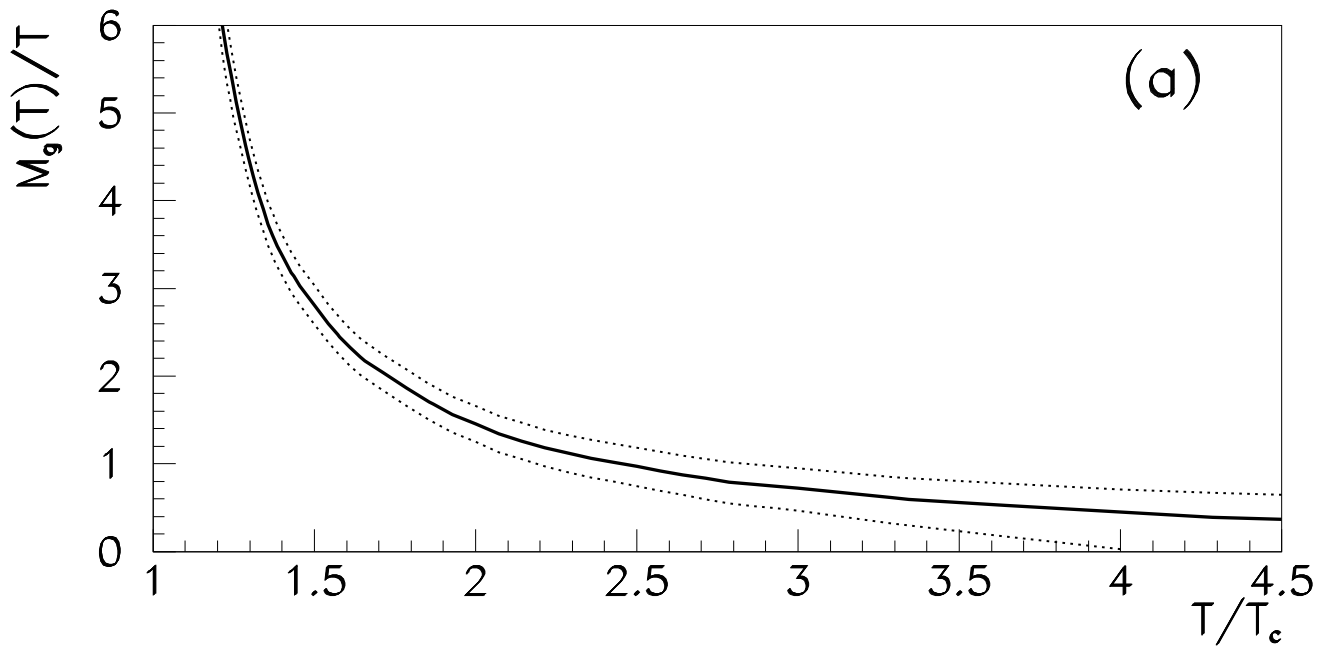
\item[Fig. 1:]
The temperature dependent effective gluon mass,
$M_g(T)$, in units of  temperature $T$ (a)
and the effective number of gluonic quasi-particle
degrees of freedom, $D(T)$ (b) in pure SU(3) gauge theory ($N_f$=0)
as a function of $T/T_c$. The horizontal line in Fig. 1b indicates $D_g=16$.

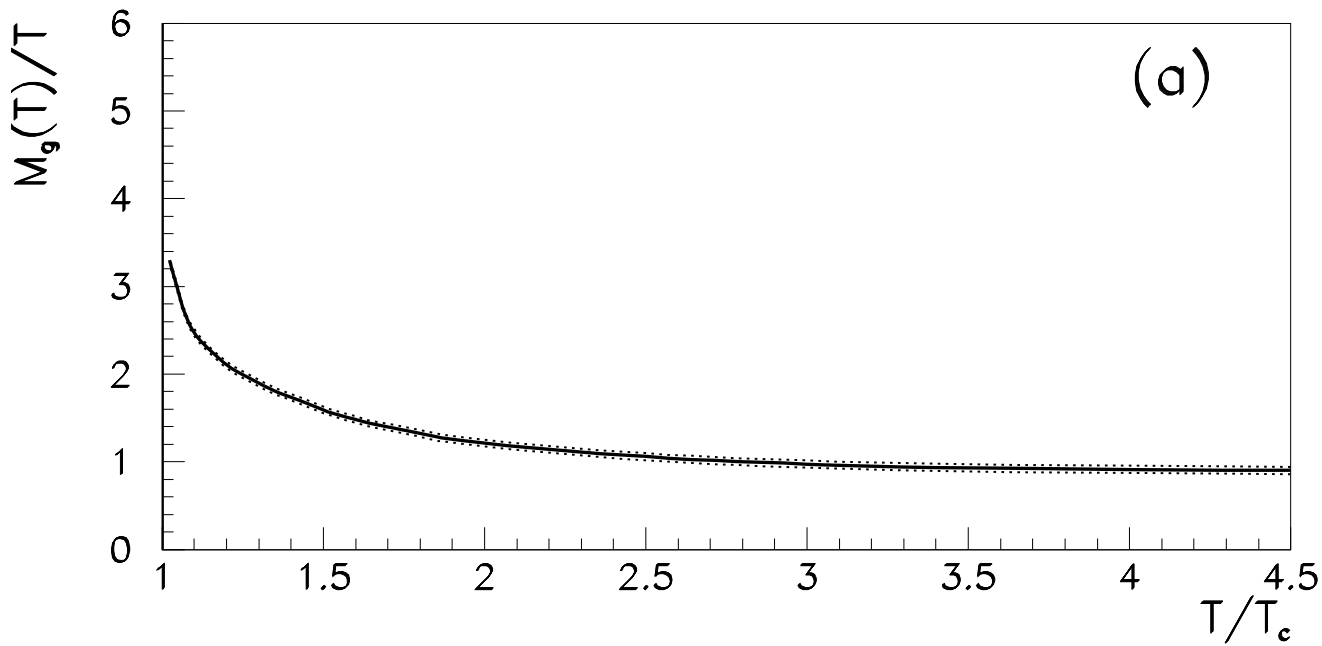
\item[Fig. 2:]
The temperature dependent effective gluon mass,
$M_g(T)$, in units of temperature $T$ (a)
and the interaction term $B(T)$
normalized by the energy density $\varepsilon(T)$ (b)
for pure SU(3) gauge theory ($N_f$=0)
as a function of $T/T_c$. For details see text.

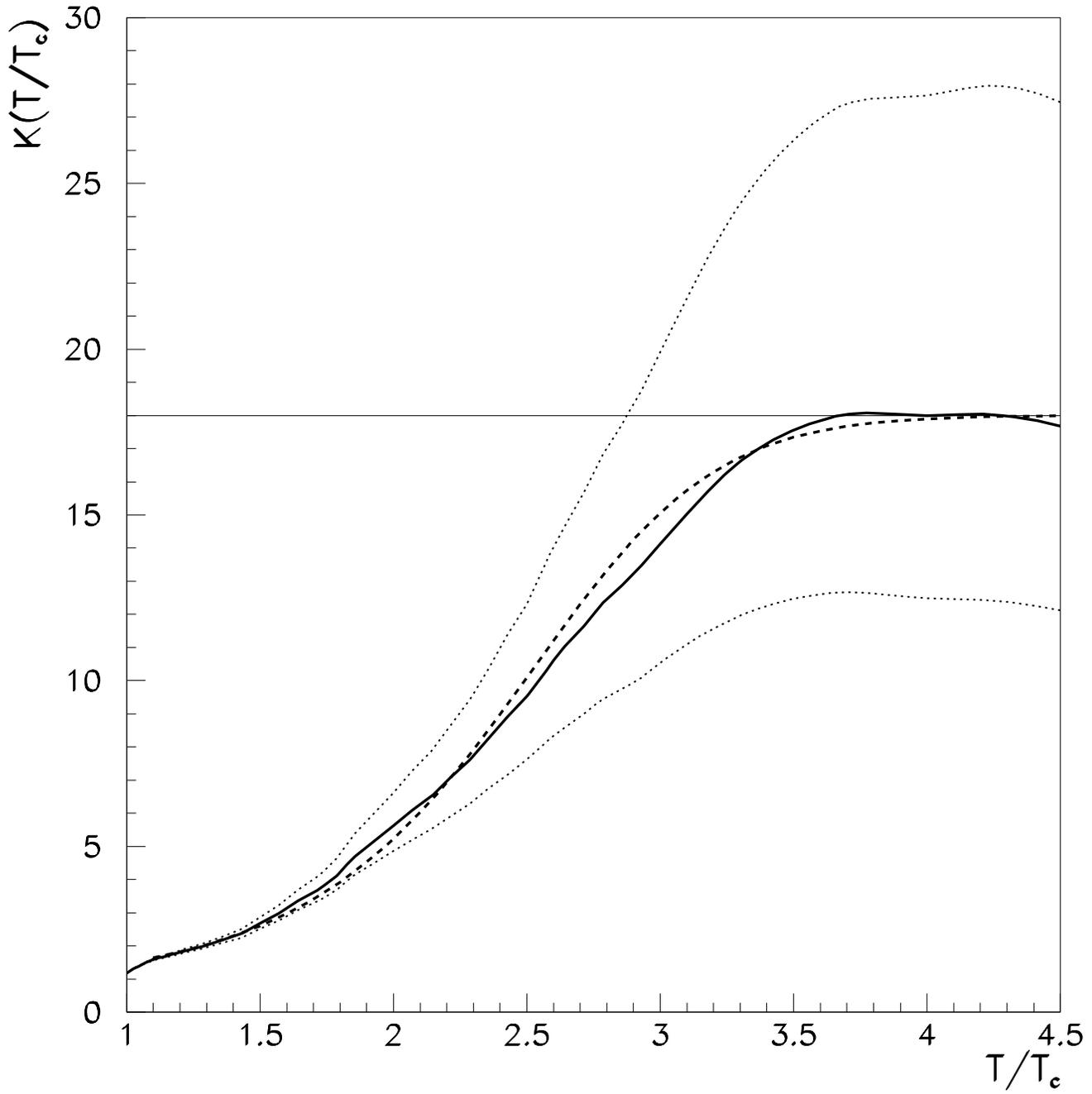
\item[Fig. 3:]
The function $K(T/T_c)$ for the effective coupling constant $g(T)$
in pure SU(3) gauge theory.
The solid line shows the mean values, the dotted lines show the 
influence of a $\pm 2\%$
systematic error of the lattice data on the function
$K(T/T_c)$. The fitted function Eq.~(\ref{kfun}) is denoted
by the dashed line, and the horizontal line at $K=18$ indicates its
asymptotic value.

\item[Fig. 4:]
Lattice data for pure SU(3) gauge theory on $\varepsilon/T^4$
and $3P/T^4$ (diamonds) \cite{Boyd95},
their reconstruction with a free
quasi-particle picture with $B(T)=0$ (dash-dotted lines),
 and including the
interaction term $B(T)$ (dashed lines).
The horizontal line at $5.24$ indicates the Stefan-Boltzmann
value for massless non-interacting particles.

\item[Fig. 5:]
The temperature dependent effective gluon mass,
$M_g(T)$, in units of  temperature $T$ (a)
and the interaction term $B(T)$
normalized by the energy density $\varepsilon(T)$ (b)
in SU(3) gauge theory for $N_f$=2 and $N_f$=4
as a function of $T/T_c$. 

\item[Fig. 6:]
The function $K(T/T_c)$ from the effective coupling constant $g(T)$
in SU(3) gauge theory $N_f=0,2,4$ 
(solid lines) and the influence of 
$\pm 2 \%$  systematic error of lattice data on the function
$K(T/T_c)$. 

\item[Fig. 7:]
SU(3) lattice data (symbols) and our model reconstruction (lines):
$N_f=2$ \cite{SU3NF2a} (a) and $N_f=4$ \cite{SU3NF4} (b).
The reconstructed quantities are denoted by dash-dotted lines
in a simple quasi-particle picture and  dashed lines if
the interaction term $B(T)$ is included.
Horizontal lines indicate the Stefan-Boltzmann
values for massless non-interacting particles, 
12.17 and 19.08, respectively.

\item[Fig. 8:]
The speed of sound $c_s^2$, calculated for $N_f=0,2,4$. The 
dash-dotted lines  denote the results $c_{s,0}^2$
from the free quasi-particle model ($B(T)=0$)
while the dashed lines with asymptotic value 
$c_s^2=0.33$  indicate the results including the interaction term
$B(T)$. Note that the differences between $N_f=$ 0, 2 and 4 are hardly 
visible.

\item[Fig. 9:]
The effective masses of quarks (dashed curves) 
and gluons (solid curves)  in SU(3) gauge
theory with $N_f=$0, 2 and 4 quark flavour.

\item[Fig. 10:]
The density ratios between massive ($n_i$) and massless
($n_i^0$) quarks and gluons in SU(3) gauge theory
for $N_f=$0, 2 and 4.

\item[Fig. 11:]
The relative gluon density 
$R_{g}=n_g/(n_g+n_q+n_{\overline q})$
for massive, $R_{g}^0$ for massless particles
in $N_f=2$ (solid lines) and in $N_f=4$
(dashed lines) QGP.

\end{description}

\vfill
\eject
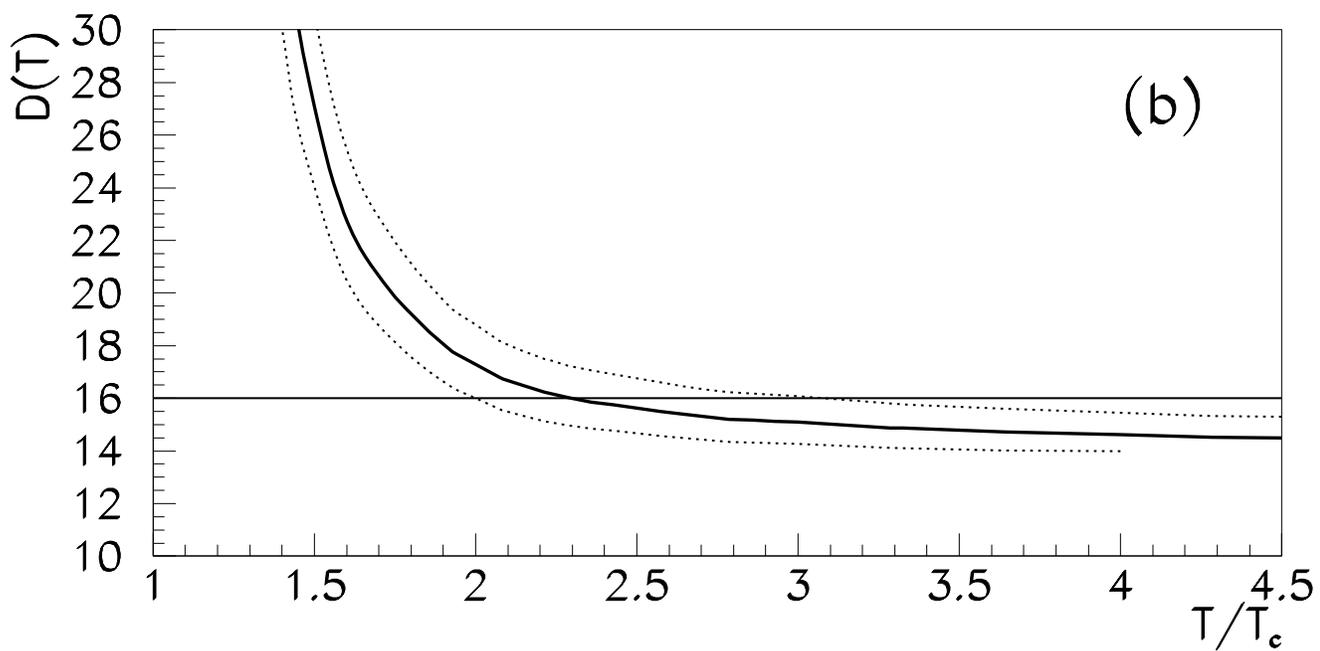
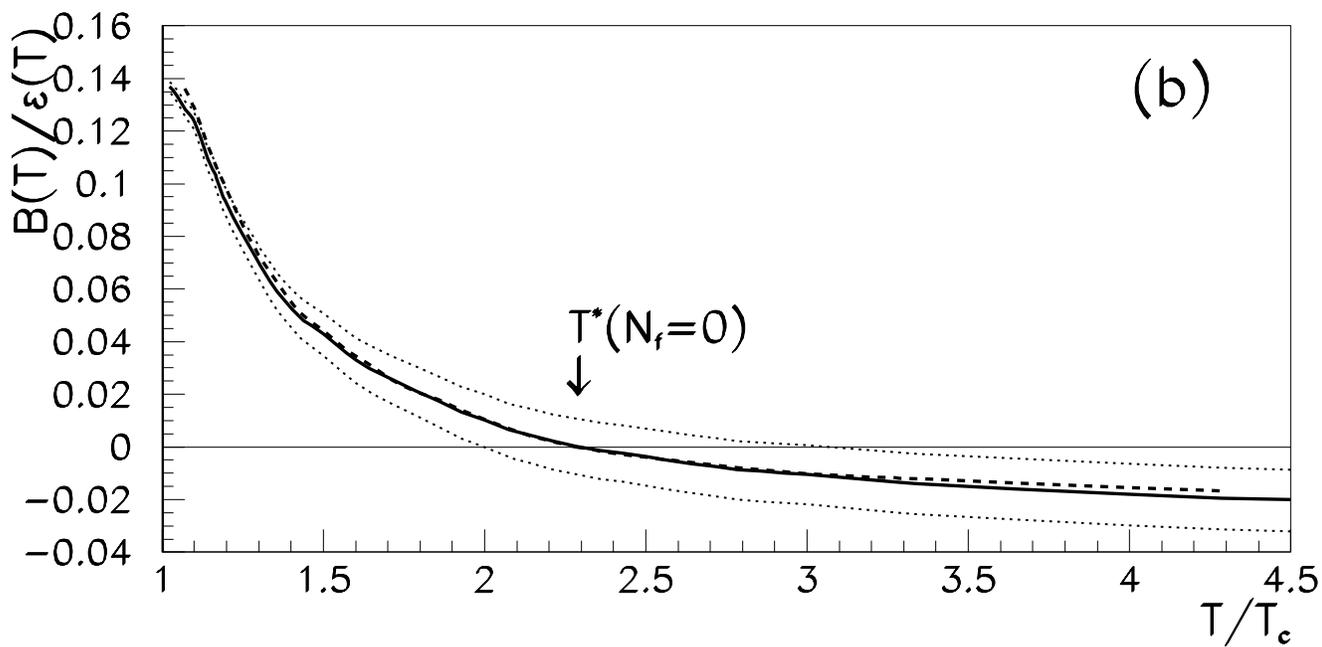
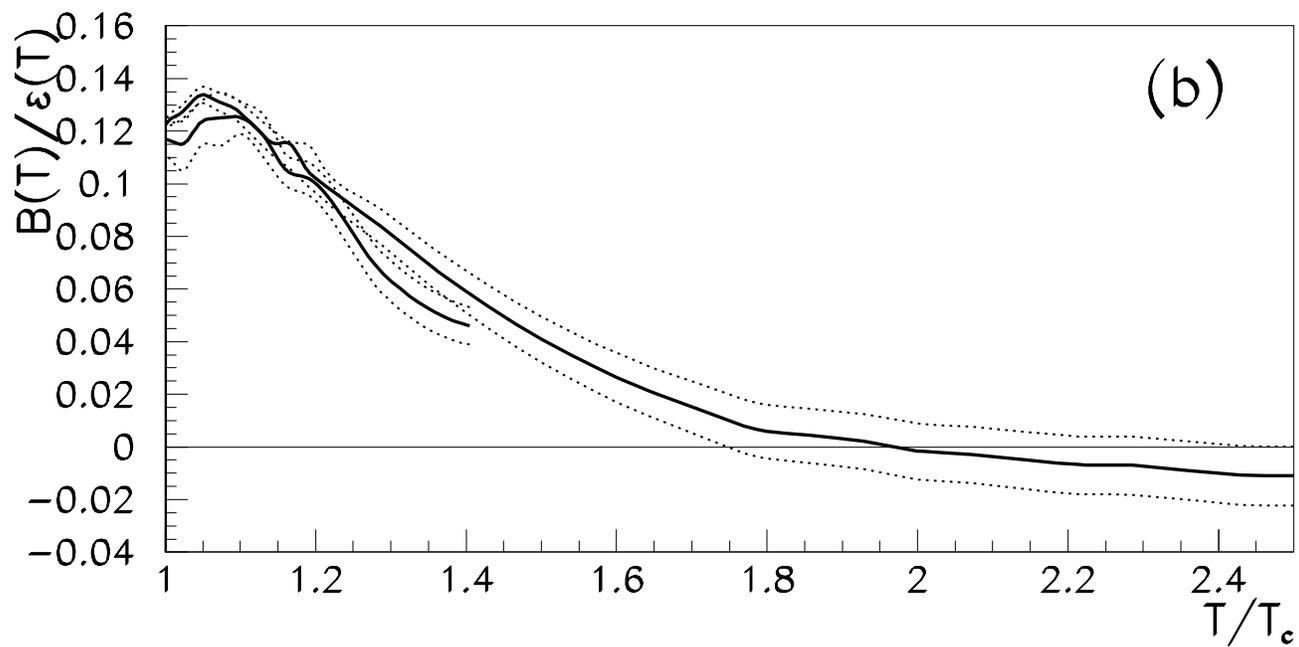
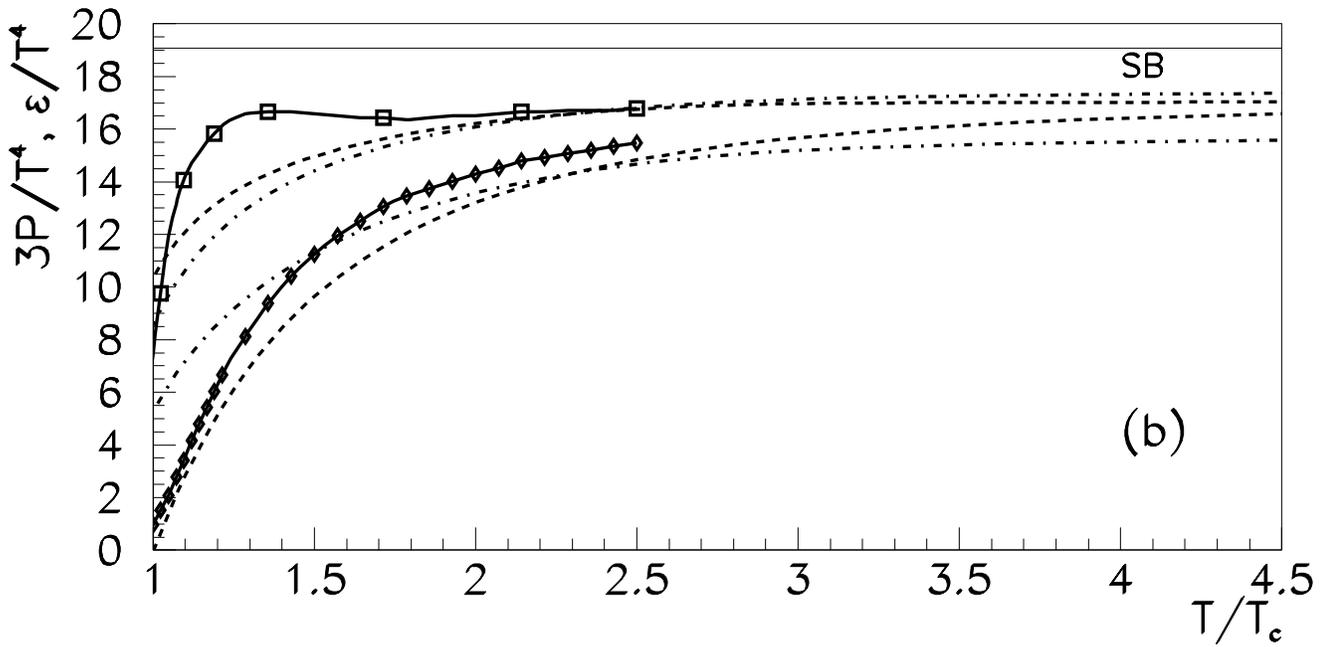
\end{document}